# Volatile loss following cooling and accretion of the Moon revealed by chromium isotopes

Paolo A. Sossi[a*], Frédéric Moynier[a, b], Kirsten van Zuilen[a,^]

[a] Institut de Physique du Globe de Paris, Université Paris Diderot, Université Sorbonne Paris Cité, CNRS UMR 7154, 1 rue Jussieu, 75238 Paris Cedex 05, France

[b] Institut Universitaire de France, Paris, France

[^] Present address: Department of Earth Sciences, Vrije Universiteit Amsterdam, 1081 HV Amsterdam, The Netherlands

* Corresponding author: sossi@ipgp.fr             .

## Abstract

Terrestrial and lunar rocks share chemical and isotopic similarities in refractory elements, suggestive of a common precursor. By contrast, the marked depletion of volatile elements in lunar rocks together with their enrichment in heavy isotopes compared to Earth's mantle suggests that the Moon underwent evaporative loss of volatiles. However, whether equilibrium prevailed during evaporation, and, if so, at what conditions (temperature, pressure and oxygen fugacity) remain unconstrained. Chromium may shed light on this question, as it has several thermodynamically-stable, oxidised gas species that can distinguish between kinetic and equilibrium regimes. Here, we present high-precision Cr isotope measurements in terrestrial and lunar rocks that reveal an enrichment in the lighter isotopes of Cr in the Moon compared to Earth's mantle by 100±40 ppm per atomic mass unit. This observation is consistent with Cr partitioning into an oxygen-rich vapour phase in equilibrium with the proto-Moon, thereby stabilising the $CrO_2$ species that is isotopically heavy compared to CrO in a lunar melt. Temperatures of 1600 – 1800 K and oxygen fugacities near the Fayalite-Magnetite-Quartz buffer are required to explain the elemental and isotopic difference of Cr between Earth's mantle and the Moon. These temperatures are far lower than modelled in the aftermath of a giant impact, implying that volatile loss did not occur contemporaneously with impact but following cooling and accretion of the Moon.

**Significance**

*Excepting volatile elements, which are strongly depleted and isotopically fractionated, the Moon has chemical and isotopic signatures that are indistinguishable from Earth's mantle. Reconciliation of these properties with Moon formation in a high-energy giant impact invokes evaporative loss of volatile elements, but at conditions that are poorly-known. Chromium isotopic fractionation is sensitive to temperature variations and liquid-gas equilibration during evaporation. We measure an isotopic difference between Earth's mantle and the Moon consistent with the loss of a Cr-bearing, oxidised vapour phase in equilibrium with the proto-Moon. Temperatures of vapour loss required are much lower than predicted by recent models, implying that volatile elements were removed from the Moon following cooling, rather than during a giant impact.*

**Body**

The volatile-depleted nature of the Moon was recognised upon analysis of the first lunar samples returned by the Apollo missions (1), yet it remains one of its most enigmatic characteristics (2, 3). The gross depletions in volatile elements in the lunar mantle relative to Earth's stand in contrast to the chemical and isotopic similarity observed for refractory elements (4–6). Although some models account for this accord by a giant impact in which the impactor had an isotopic composition identical to that of Earth (7), this explanation cannot account for the impoverishment of Cr, Mn and V in the lunar mantle. The depletion in these elements in both the lunar and terrestrial mantles relative to other solar system bodies uniquely resulted from the high pressure under which Earth's core formed (8–10), and constitutes strong evidence for the derivation of the Moon from Earth's mantle (4).

Within this genetic framework, however, the thermochemical conditions that could engender similarity in refractory lithophile element abundances yet strong volatile element depletion in the Moon remain poorly understood. Based on the observation that lunar mare basalts are depleted only in elements more volatile than Li, *e.g.* (11), temperatures of ≈1100 K for lunar volatile loss have been proposed (2, 12), reflecting the nebular half-condensation temperature of Li. However, this conclusion is contingent upon the assumption that element volatility during condensation of the solar nebula is transferable to that for evaporation of lunar silicates. Equilibrium thermodynamic calculations provide grounds to reject this assumption, as the vapour calculated to be in equilibrium with a model lunar mantle composition has oxygen fugacities orders of magnitude higher (13, 14) than in the solar nebula. These thermodynamic variables affect element volatility, and new dynamical simulations invoke higher temperatures (>2500 K) to simultaneously re-produce lunar volatile depletion in the Earth-Moon disk, and equilibrate refractory elements between the two bodies (14, 15).

Stable isotopes can shed light on these discrepant temperature estimates because, at equilibrium, their fractionation between two phases (*e.g.,* liquid and gas) is proportional to $1/T^2$. The enrichment in the heavier isotopes of Zn, K, Rb, Ga and Cl in lunar rocks relative to Earth's mantle attests to their evaporation and subsequent escape from the Moon (16–21). Whether this loss occurred during local magmatic degassing (20, 22), a magma ocean phase (17, 23, 24) or following a giant impact (1, 16) remains debated. Determining the conditions of volatile depletion therefore has the potential to distinguish between these scenarios. However, it has been hitherto impossible to quantify evaporation temperatures using the isotope compositions of Zn, K, Ga and Rb, because their vapour species are monatomic gases, *e.g.,* $Zn^0$ (25), which favour the lighter isotopes with respect to the condensed phase, *e.g.,* ZnO (26). As such, the direction of isotopic fractionation is the same, be it at equilibrium or during kinetic vapour loss into a vacuum, for which the fractionation factor is temperature-independent. Chromium represents a special case; although $Cr^0_{(g)}$ is stable in the solar nebula, unlike the aforementioned elements, it also has several oxidised gas species, namely, $CrO_{(g)}$, $CrO_{2(g)}$ and $CrO_{3(g)}$ (27). Therefore, under the more oxidised conditions that typify evaporation of planetary mantles (13), Cr becomes more volatile, whereas other moderately volatile elements (*e.g.,* Rb, Zn) become less so. As such, evaporative loss of Cr is possible under these conditions (11) and could engender light isotope enrichment in the residue.

In order to evaluate whether there are any Cr isotope differences between Earth and the Moon, 17 spinifex-textured komatiites, whose compositions approximate those of liquids, were sourced from five different cratons (28). These samples constitute records of Earth's mantle composition in the Archean, and together with three modern peridotites, they are used to estimate its Cr isotope composition through time and space. Estimates of the Moon's composition were garnered by analysis of six Mg Suite samples, a set of cogenetic cumulate rocks that record the early stages of the Moon's differentiation (29), in addition to five lunar mare basalts and lunar green glass. The Cr isotopic data are reported as $\delta^{53}$Cr, the per mille deviation of the $^{53}$Cr/$^{52}$Cr ratio of the sample from the NIST SRM 979 standard, together with its associated 2SD uncertainty throughout unless otherwise specified.

**Table 1.** Chromium isotopic composition, chromium content and MgO content of samples.

| Earth | Rock type | $\delta^{53}$Cr (‰) | 2×s.d. | n | Cr (ppm) | MgO (wt. %) |
|---|---|---|---|---|---|---|
| 49J | Komatiite | -0.065 | 0.041 | 3 | 3190 | 32.16 |
| B-R1 | Komatiite | -0.155 | 0.035 | 3 | 2650 | 27.72 |
| 179/751 | Komatiite | -0.122 | 0.066 | 3 | 3240 | 23.54 |
| 176/723 | Komatiite | -0.122 | 0.048 | 3 | 2450 | 31.13 |
| SD5/354.5 | Komatiite | -0.121 | 0.017 | 2 | 3050 | 25.72 |
| 331/783 | Komatiite | -0.082 | 0.039 | 3 | 2894 | 26.69 |
| 331/777A | Komatiite | -0.114 | 0.012 | 2 | 2717 | 26.26 |
| 422/94 | Komatiite | -0.147 | 0.035 | 2 | 3464 | 22.41 |
| 422/95 | Komatiite | -0.168 | 0.059 | 2 | 3260 | 23.45 |
| 422/84 | Komatiite | -0.112 | 0.004 | 2 | 2902 | 30.32 |
| RL-12-1 | Kom. Basalt | -0.111 | 0.010 | 2 | 1351 | 13.68 |
| 331/790 | Px. Cumulate | -0.132 | 0.006 | 2 | 3371 | 16.05 |
| 422/96 | Komatiite | -0.097 | 0.001 | 2 | 3060 | 28.71 |
| SD6/400 | Komatiite | -0.098 | 0.004 | 2 | 2428 | 27.99 |
| 422/96a | Komatiite | -0.085 | 0.001 | 2 | 3060 | 28.71 |
| 331/948 | Komatiite | -0.104 | 0.004 | 2 | 2915 | 23.54 |
| 331/779 | Komatiite | -0.097 | 0.004 | 2 | 2706 | 26.76 |
| DTS-1 | Dunite | -0.127 | 0.007 | 2 | 3990 | 49.59 |
| PCC-1 | Harzburgite | -0.083 | 0.004 | 2 | 2730 | 43.43 |
| DTS-1a | Dunite | -0.123 | 0.003 | 2 | 3990 | 49.59 |
| *Moon* | | | | | | |
| 15445 | Norite | -0.082 | 0.007 | 2 | 1635 | 10.20 |
| 15455 | Norite | -0.116 | 0.009 | 2 | 810 | 10.90 |
| 76535 | Troctolite | -0.253 | 0.005 | 2 | 753 | 19.09 |
| 78235 | Norite | -0.140 | 0.003 | 2 | 1500 | 11.76 |
| 78328 | Norite | -0.138 | 0.006 | 2 | 1500 | 11.76 |
| 72415 | Dunite | -0.297 | 0.011 | 2 | 2414 | 43.61 |
| 15555 | Low Ti | -0.283 | 0.004 | 2 | 4290 | 11.10 |
| 12002 | Low Ti | -0.234 | 0.009 | 2 | 5949 | 14.82 |
| 15426 | Green Glass | -0.186 | 0.009 | 2 | 3557 | 17.50 |
| 10003 | High Ti | -0.307 | 0.007 | 2 | 1583 | 7.10 |
| 70135 | High Ti | -0.196 | 0.012 | 2 | 3921 | 9.30 |
| 10057 | High Ti | -0.243 | 0.002 | 2 | 2303 | 7.60 |

*n denotes the number of replicates on the MC-ICP-MS.

Twenty terrestrial ultramafic samples have $\delta^{53}$Cr between -0.17±0.06 ‰ (sample 422/95) and -0.07±0.04 ‰ (sample 49J), with an average of -0.11±0.05‰ (Table 1), identical to the average of samples used to define bulk silicate Earth (BSE), -0.13±0.10‰ (30) but with less scatter. The lack of isotopic variation despite differences in *i)* different emplacement ages (3.5 to 2.7 Ga for komatiites; present-day for peridotites), *ii)* komatiite source fertility and *iii)* lithology (peridotites *vs.* komatiites) attests to the homogeneity of Cr isotopes in convecting mantle from 3.5 Ga to the present-day. The Cr isotope compositions for the five mare basalts range from -0.31±0.01 ‰ in sample 10003 to -0.20±0.01 ‰ in sample 70135, both of which are high Ti basalts and span the range of the two low Ti basalts (Table 1). A previous study found overlapping Cr isotope compositions in Low-Ti and High-Ti basalts, with an average of -0.22±0.10 ‰ (31). As a whole, lunar mare basalts tend to lighter $\delta^{53}$Cr with decreasing Cr (Fig. 1a) and MgO content (Fig. 1b). By contrast, the Mg Suite cumulates record an increase of $\delta^{53}$Cr with falling MgO (Fig. 1b). The olivine-rich samples dunite 72415 and troctolite 76535 have $\delta^{53}$Cr within the range of lunar mare basalts (-0.30±0.01‰ and -0.25±0.01‰, respectively), while the more evolved norites (orthopyroxene-plagioclase) have distinctly heavier compositions, up to -0.08±0.01‰. Green glass clods from soil sample 15426 have $\delta^{53}$Cr = -0.19±0.01‰.

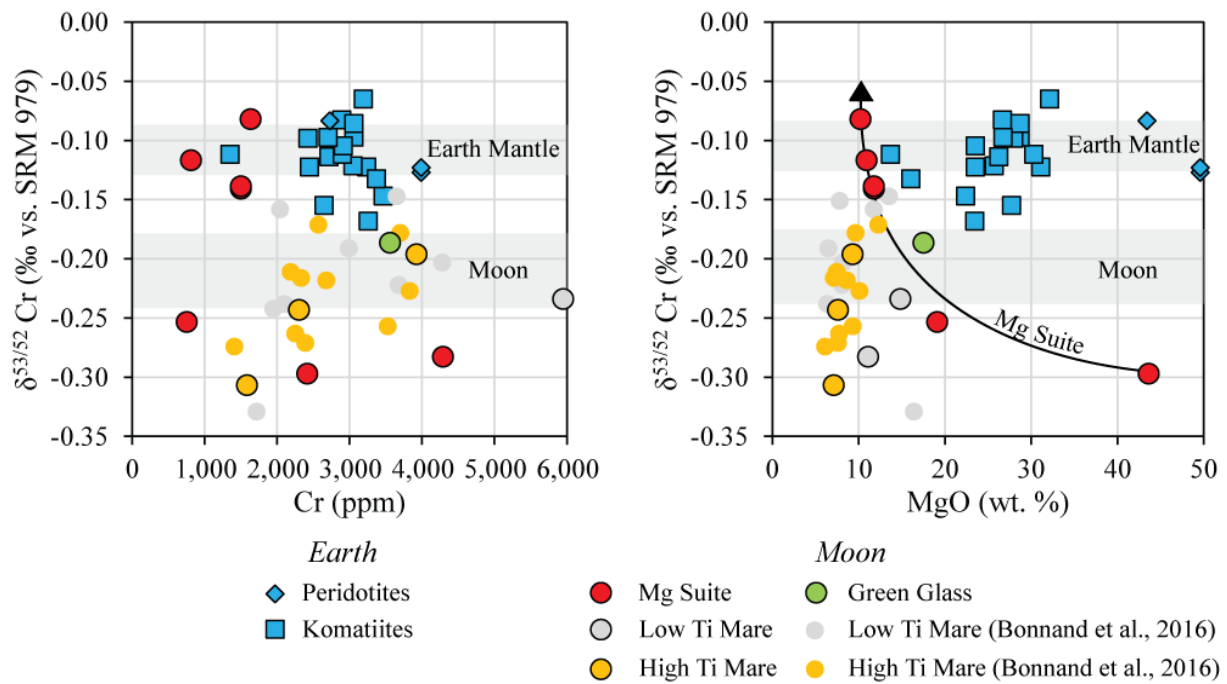

**Fig. 1.** The chromium isotope composition of lunar samples; Mg Suite = red (*n* = 6), Green glass = green (*n* = 1), High Ti = yellow (*n* = 13), Low Ti = grey (*n* = 11), smaller circles show the data of (30); and terrestrial komatiites (blue squares, *n* = 17) and peridotites (blue diamonds, *n* = 3), expressed as $\delta^{53}$Cr, as a function of their a) chromium content and b) MgO content. Also shown are a trend line for the Mg Suite (black line), and estimates for the bulk composition of bulk silicate Earth and the Moon (grey fields).

**Mechanics of Cr isotope fractionation during magmatic processes**

Chromium isotope fractionation arises from the decoupling of $Cr^{2+}$ and $Cr^{3+}$ in magmatic phases. Their relative abundance in silicate liquids is dependent on oxygen fugacity, $fO_2$, according to the homogeneous equilibrium:

$$CrO_{1.5} = CrO + ¼O_2 \quad (1)$$

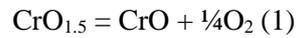

The positive entropy change of reaction (1) stabilises chromous oxide to high temperatures (32). Terrestrial and lunar magmas have $fO_2$ near the Fayalite-Magnetite-Quartz (FMQ) and about 1 log unit below the Iron-Wüstite (IW) buffers, respectively (33). Under these conditions with $\log K_{(1)} \approx 1.9$ at 1400°C, $Cr^{2+}/\sum Cr$ is 0.32 and 0.91 for terrestrial and lunar magmatic liquids, respectively (34).

In (ultra)mafic magmas, olivine, pyroxene and chromite can leverage Cr isotope fractionation in the liquid from which they crystallise. The precipitation of chromite, $(Fe,Mg)Cr^{3+}_2O_4$, is favoured by increasing $fO_2$ and falling temperature that increase $aCrO_{1.5}$ in the liquid (35). Though chromite saturation occurs with olivine in some komatiitic magmas, its effect on the Cr isotope composition of the samples measured is negligible owing to the *i)* high temperatures, >1450°C, (36) *ii)* the fact that they represent quenched liquids and have not undergone chromite accumulation or fractionation and *iii)* the high $fO_2$ and hence $Cr^{3+}$ content of terrestrial melts.

Conversely, the Cr isotope variation observed in lunar magmas is larger, up to 0.20 ‰. Lunar mare basalt parent magmas have Mg# ≈ 0.45, constrained by Fo$_{72}$ and a $K_D{}^{Fe-Mg}_{ol-melt}$ = 0.3 (37). These magmas have 1 bar liquidi of ≈1200-1250°C, at which point olivine and chromite crystallise in tandem (38). The Cr isotopic fractionation between $Cr^{3+}$-bearing chromite and a lunar silicate melt

(assuming it is equivalent to $Cr^{2+}$-bearing forsterite) with $Cr^{2+}/\sum Cr = 0.91$ is $0.35 \times 10^6/T^2$ (39), or $\Delta^{53}Cr_{chromite-melt} = +0.16$ ‰ at 1200°C, which is identical to that empirically determined by (31).

The cumulate samples of the Mg Suite, whose compositions are complementary to those of liquids, have a more magnesian parent magma (≈13 wt. % MgO; 40) than those of mare basalts, with a higher liquidus (1300°C), at which only olivine is stable. Indeed, the Cr budget in 72415 dunite and 76535 troctolite was entirely hosted in olivine (41), into which both $Cr^{2+}$ and $Cr^{3+}$ are incorporated sub-equally. Thus, minimal isotopic fractionation occurs during crystallisation of 72415 and 76535, meaning their Cr isotope composition reflects that of the Mg Suite parent magma. Relative to $Cr^{2+}$, trivalent Cr is ~ten times more compatible in orthopyroxene (35), such that even a lunar melt with ≈10% $Cr^{3+}$ will crystallise orthopyroxene with $Cr^{2+}/\sum Cr \approx 0.45$. Adopting $\Delta^{53}Cr_{Cr3+-Cr2+} = +0.35 \times 10^6/T^2$ (39), weighted for the difference in $Cr^{2+}/\sum Cr$ between orthopyroxene and melt, predicts $\Delta^{53}Cr_{opx-melt} = +0.10$ ‰ at 1200°C (SI Appendix, Fig. S1). This results in orthopyroxene-plagioclase cumulates, represented by norites 78235/8 and the more evolved 15455, with $\delta^{53}Cr$ 0.10 ‰ heavier than the Mg Suite parent magma (-0.25 ‰), in agreement with the measured values (Table 1). Norite 15445 is mineralogically similar to 15455, except it has twice the amount of Cr (1635 *vs.* 810 ppm) and a correspondingly heavier $\delta^{53}Cr$ (-0.08±0.01 ‰ *vs.* -0.12±0.01 ‰), consistent with 0.25 % chromite accumulation.

**Composition of the bulk silicate Earth and Moon**

Owing to the potential for chromite to engender isotopic fractionation, especially in lunar magmas where the disparity in valence state of Cr in the melt ($Cr^{2+}$) and in chromite ($Cr^{3+}$) is large, quenched liquids or direct mantle samples are the best indicators of mantle composition.

In determining the Cr isotope composition of the BSE, data for ultramafic rocks from this study and the literature (30, 42, 43) were filtered for i) lithologies that represent mantle-derived rocks (excluding cumulates such as wehrlites and hornblendites) and ii) fertile, unmetasomatised peridotites with Cr contents of 2520±630 ppm, Mg# (molar Mg/(Mg+Fe)) between 0.885 and 0.910, and >2 wt. % $Al_2O_3$. The remaining 42 samples with mean $\delta^{53}Cr$ = -0.11±0.11 ‰ were then filtered for statistical outliers at the 95% confidence level using Grubb's test, leaving 36 samples with $\delta^{53}$Cr = -0.11±0.06 ‰ whose population is normally-distributed (SI Appendix, Table S9) and has a 2×standard error of the mean (2SE) of ±0.02 ‰.

In order to quantify the degree of mineral accumulation in lunar mare basalts, bulk rock Mg# was compared to that expected for a liquid in equilibrium with the highest Mg# olivine (or orthopyroxene in olivine-free basalts) in the sample, given $K_{D\ ol,opx-melt}^{Fe-Mg} \approx 0.3$ (44). Samples were rejected from the derivation of the bulk silicate Moon value if they had experienced *i)* significant (> 0.1) olivine accumulation or fractionation (0.4 < Mg# < 0.5 were kept) and *ii)* chromite accumulation (high Cr and $\delta^{53}$Cr) or fractionation (low Cr and $\delta^{53}$Cr). This exercise left 17 measurements of 15 lunar samples from (31) and this study, with average $\delta^{53}$Cr = -0.21±0.06 ‰ and 2SE of ±0.03 ‰ (SI Appendix, Table S11; no outliers were found).

A two-tailed student's t-test yielded a t-statistic of 13.36 relative to a critical value of 2.00, illustrating that the $\delta^{53}$Cr of the Earth (-0.11±0.02 ‰, 2SE, $n$ = 36) and Moon (-0.21±0.03 ‰, 2SE, $n$ = 17) are clearly statistically resolvable at $\Delta^{53}$Cr$_{Moon-Earth}$ = -0.10±0.04‰.

Estimates for the Cr content of the lunar mantle, based on Fe/Cr and on Cr/V correlations in lunar rocks give 2200 ppm and 2500 ppm (8), respectively. The latter estimate is a maximum because it assumes V is refractory, despite the fact that Al/V ratios increase from 159 in CI chondrites to 182 in CV chondrites (9). If V is slightly depleted in the Moon (by a factor of 159/182), then its Cr content decreases to 2180 ppm. A value of 2000 ppm is calculated by combining the ≈3500 ppm Cr in lunar green glass beads with chromium partitioning during partial melting of an olivine-dominated lunar mantle source (9). An average of these estimates gives 2125±110 ppm Cr in the Moon, which is therefore depleted with respect to the Earth's mantle (2520±250 ppm) by 16±10%.

**Isotopic fractionation of chromium during planetary formation**

By adopting the Earth's mantle as the Moon's progenitor (4, 40), either core formation or volatile loss can simultaneously decrease Cr abundance and shift its isotopic composition.

Geophysical results from the GRAIL mission have shown that the core comprises 1% of the Moon's mass (45). At 55 kbar, the pressure at its centre, the partition coefficient of Cr into Fe-Ni metal is very near unity at IW-2 (10), and, as such, the lunar core would only contribute to a minor (≈1%) depletion of Cr, attested to by the paucity of solutions in lunar core formation models that fit observed Cr depletion (46). Furthermore, *ab-initio* calculations show that the isotopic composition of metallic Cr is slightly lighter than $Cr^{2+}$ in the M-sites of olivine (39) meaning metal segregation would result in isotopically heavy silicates, contrary to observations. It is therefore unlikely that core formation can account for the Cr depletion or its light isotopic composition in the lunar mantle.

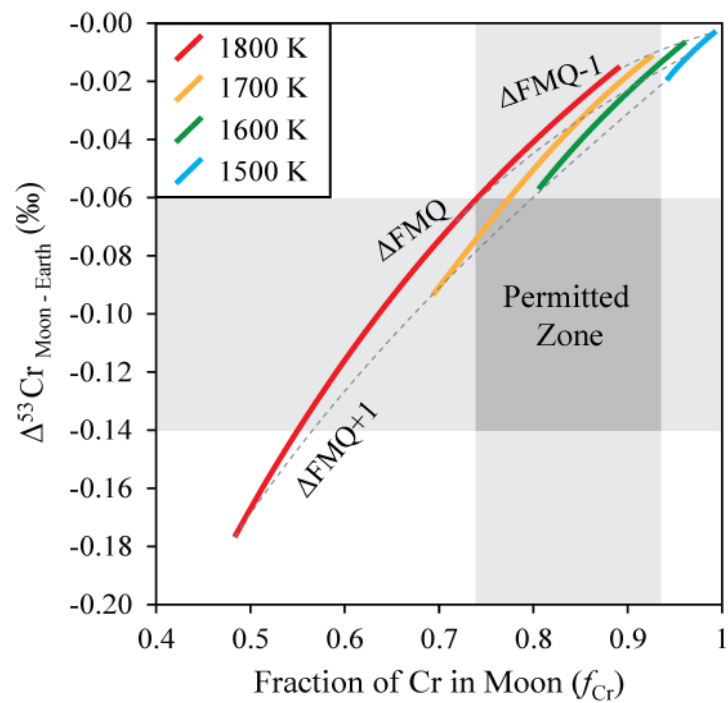

**Fig. 2.** The fraction of Cr remaining in the bulk silicate Moon relative to the bulk silicate Earth (BSE), $f_{Cr}$, as a function of the $\delta^{53/52}Cr$ isotopic difference between the Moon and BSE, $\Delta^{53}Cr_{Moon-Earth}$. The curves correspond to values of $f_{Cr}$ and $\Delta^{53}Cr_{Moon-Earth}$ calculated by liquid-vapour equilibrium between $CrO_2(g)$ and $CrO(l)$ at various temperatures (1500 K, blue; 1600 K, green; 1700 K, yellow and 1800 K, red) and oxygen fugacities (dashed lines, ΔFMQ+1, ΔFMQ and ΔFMQ-1). Values of $f_{Cr}$ are calculated from equation (2) and depend on temperature, pressure and $fO_2$, where the total pressure is that in the gas above a BSE composition (13). Values of $\Delta^{53}Cr_{Moon-Earth}$ are given by equation (4). The dark grey field demarcates the permissible range of $f_{Cr}$ and $\Delta^{53}Cr_{Moon-Earth}$ defined by sample data; curves passing through this field satisfy both constraints.

Vaporisation of silicate material of a bulk lunar composition gives rise to $fO_2$ of IW+2.5 at 1800K (13). The relatively oxidising vapour, near the FMQ buffer, evolved upon evaporation of silicate material, be it at high- (1827-1970K, olivine; 47) or low temperature (1396-1499K, lunar mare basalt 12002; 48) is evidenced by experimental Knudsen Effusion Mass Spectrometry studies. Gas phase equilibria among the four Cr oxide species (27; SI Appendix, Fig. S2) show that $CrO_{2(g)}$ is stable for all $fO_2 >$ IW. Considering the predominance of $Cr^{2+}$ in a lunar silicate liquid (34), the appropriate evaporation equation is:

$$CrO(l) + \tfrac{1}{2}O_2 = CrO_2(g). \quad (2)$$

Increasing oxygen fugacity increases the volatility of Cr (eq. 2), in contrast to Zn, Rb, Ga or K (25). The partial pressure of $CrO_{2(g)}$ is calculated in a vapour in equilibrium with a BSE-like silicate melt composition (13) as a function of temperature and $fO_2$ (Fig. 2). Given an ideal gas, significant (>1%) $CrO_{2(g)}$ in the vapour only occurs above 1500 K (Fig. 2), assuming a CrO activity coefficient in silicate melts of ≈3 (34).

Whether gas-liquid exchange between $CrO_{2(g)}$ and $CrO_{(l)}$ can drive the condensed phase to isotopically light compositions requires knowledge of the force constants of Cr in both $CrO_{(l)}$ and $CrO_{2(g)}$, neither of which are available. To remedy this, we calculated the reduced partition function ratio of $^{53}Cr/^{52}Cr$ substitution in the diatomic molecule $CrO_{(g)}$ in the harmonic approximation (49) given the Cr-O vibrational frequency of 864 cm$^{-1}$ (50). Then, bond valence theory was used in the electrostatic description of bonding with Born-Landé potentials (51) to predict the mean bond strength of $CrO_{(g)}$, $CrO_{2(g)}$ and $CrO_{3(g)}$, giving $10^3\ln\beta^{53/52}Cr$ of 0.28, 0.57 and 0.89×$10^6/T^2$, respectively (SI Appendix, Table S12, Fig. S3). This approach yields force constants that agree with spectroscopic or experimental determinations to within 50% (×$10^6/T^2$). Using the $10^3\ln\beta^{53/52}Cr$ of the $Cr_2SiO_4$ component in olivine, 0.26×$10^6/T^2$, as an analogue for $CrO_{(l)}$ (39), the $^{53}Cr/^{52}Cr$ isotope fractionation factor for eq. 2 may be written:

$$\Delta^{53}Cr_{CrO(l)-CrO_2(g)} = -0.31 \pm 0.16 \times \frac{10^6}{T^2} \text{ (‰)} \quad (3)$$

The composition of the lunar mantle in equilibrium with a $CrO_2$-bearing gas phase can be calculated by mass balance; in which:

$$\delta^{53}Cr_{vapour} = \frac{\delta^{53}Cr_{system} - \delta^{53}Cr_{silicate\ moon} f^{Cr}_{silicate\ moon}}{(1 - f^{Cr}_{silicate\ moon})} \quad (4)$$

The $\delta^{53}Cr_{system}$ is present-day Earth's mantle (-0.11±0.02 ‰), $\delta^{53}Cr_{silicate\ moon}$ is -0.21±0.03 ‰, and $f^{Cr}_{silicate\ moon}$ is 0.84±0.10, yielding $\delta^{53}Cr_{vapour}$ = +0.41±0.33 ‰ or (+2.00/-0.45) ‰ considering uncertainty in the Cr depletion in the lunar mantle. Solving equation (3) for temperature and propagating uncertainties yields ≈ 700 (+1100/-300) K. Importantly, both thermodynamic and isotopic constraints converge at temperatures between 1600 – 1800 K and $fO_2$ between FMQ and FMQ+1 (Fig. 2).

**Implications for the origin of the Moon**

Relative to terrestrial ultramafic rocks, the uniformly light Cr isotope composition of samples most representative of the lunar mantle, independent of whether the samples are extrusive (*e.g.,* green glass) or intrusive (Mg Suite) supports a global-scale vapour-loss event, manifest in the depletion of moderately volatile elements (2, 4, 11). In contrast with recent numerical giant impact models that predict temperatures in excess of 4000 K (15, 52), the temperature of volatile loss must have been <1800 K to produce measurable equilibrium isotope fractionation of Cr (Fig. 2). Moreover, Zn, Ga and Cl exhibit variable isotopic compositions in different lithologies, suggesting that lunar volatile loss occurred during magma ocean crystallisation, at temperatures not higher than the silicate liquidus (23, 24). If volatile depletion did occur as a direct result of the impact, evaporation at the highest temperatures did not leave any stable isotopic trace. Instead, isotopic fractionation and loss must have occurred after the Earth-Moon system had fallen from its peak temperatures.

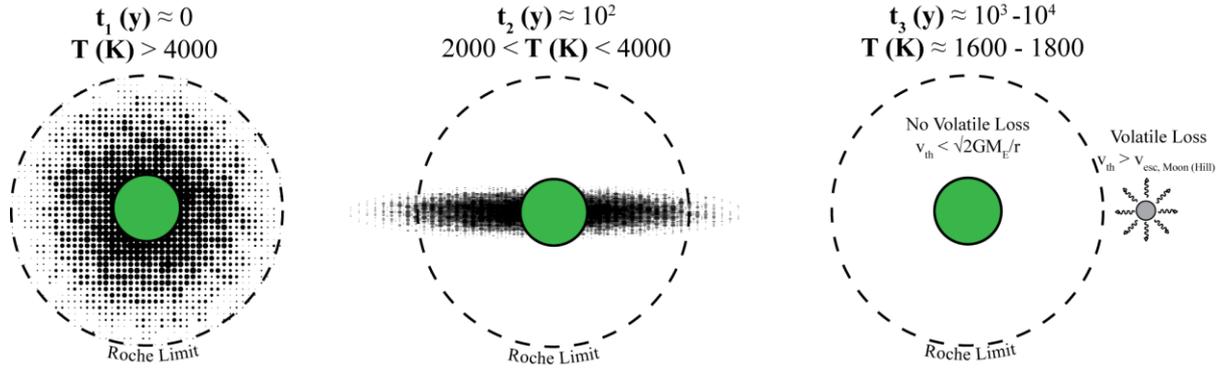

**Fig. 3.** A schematic illustration of the timing and conditions of events following a giant impact. a) $t_1$, 0 yr. A hot (> 4000 K) gas- and dust-rich vapour envelope surrounds the proto-Earth within its Roche limit. At 4000 K, >99.99 % of the volatile species of molar mass 0.052 kg/mol have $v_{th}$ < 3.8 km/s, significantly lower than the 11.2 km/s $v_{esc}$ of Earth, preventing any volatile escape of Cr. b) As the disk cools, the vapour disk rotates, settles, and spreads along the midplane, exceeding the Roche limit over a period of ≈100 years. c) The material outside the Roche limit is able to condense and accrete to form the Moon. At this point the temperature has fallen to 1600 – 1800 K, and the remaining gas particles have a Maxwell-Boltzmann distribution of velocities, 2-4% of which exceed the escape velocity of the Moon at a given time, which is calculated to occur at the Hill radius at 3 $R_E$ to be 1.56 km/s and are lost over the timescale of years.

This is difficult to reconcile with the supposition that loss of a volatile-laden atmosphere is facilitated by higher temperatures (*T*), which impart higher velocities on its constituent particles ($v_{th}$) of mass *m*, such that more exceed the escape velocity of a body ($v_{esc}$), a condition that is quantified by the escape parameter (53):

$$\lambda_{esc} = \frac{mv_{esc}^2}{2k_BT}. \quad (5)$$

Where a value of ~3 marks the transition between the Jeans (>3) and Hydrodynamic (<3) escape regimes. Given that the temperature of the atmosphere decays over time, $t_{1/2}$ ≈100 yr (54, 55), the likelihood of gas escape should be highest immediately after the giant impact (eq. 5). However, the $v_{esc}$ of a body is proportional to its mass. Prior to the accretion of the Moon, the bulk of the material lies within the Roche limit (Fig. 3; 56), where the combined masses of the Earth and Moon result in high $v_{esc}$ (11.2 km/s), thereby rendering volatile escape difficult, especially in the presence of an $O_2$- and SiO-dominated vapour (57). Indeed, at 4000 K, 99.99 % of molecules (*m* = 0.052 kg/mol) in the vapour have $v_{th}$ < 3.8 km/s. However, as the disk cools, material migrates beyond the Roche limit by viscous spreading and begins to form the proto-Moon within 1000 yr of the initial collision (55). Over this time, the disk has cooled sufficiently to temperatures (54) that are consistent with estimates based on Cr isotopes (1600-1800 K; Fig. 3). The formation of the Moon facilitates volatile escape due to the decrease in escape velocity compared to that of Earth. A particle is lost from the Moon's atmosphere when it is trapped by the gravity field of Earth; a boundary called the Hill sphere, and is:

$$v_{esc,\ Moon\ (Hill)} = \sqrt{2GM_{Moon}(\frac{1}{r} - \frac{1}{r_H})}, (6)$$

Where $G$ is the gravitational constant, and $M$ and $r$ are the Moon's mass and radius, and $r_H$ is the Hill sphere radius ($r_H = a(M_M/3M_E)^{1/3}$), where a is the distance between $M_M$ and $M_E$ (taken to be the Roche limit, 3 $R_E$). Gaseous molecules in the atmosphere surrounding the Moon have a Maxwell-Boltzmann distribution of velocities for which 2% (1600 K) to 4% (1800 K) exceed the Hill escape velocity of the Moon. In this scenario, $\lambda_{esc} \approx 4\text{-}5$, meaning loss of Cr from the lunar atmosphere is expected to occur *via* Jeans escape. Continued outward migration of the Moon beyond 3 $R_E$ increases $v_{esc}$ (eq. 6) and hence $\lambda_{esc}$. If Jeans escape controlled the lunar abundances and stable isotope compositions of the moderately volatile elements their fractionation should depend on $(m_1/m_2)^{0.5}$, causing them to be isotopically heavy in the Moon. This is difficult to uniquely assess for elements like Zn or K (17, 18) because a combination of both equilibrium- and kinetic isotope fractionation induced by Jeans escape could have produced the heavy isotopic compositions observed. Conversely, that Cr isotopes are lighter in the Moon relative to Earth implies volatile loss *via* Jeans escape was either unimportant (*i.e.*, loss was driven by another process), or went to completion resulting in quantitative loss of the Cr (and lighter elements) present in the lunar atmosphere over the timescales for cooling of the Moon.

**Materials and Methods**

*Chromium double spike*

A $^{50}$Cr-$^{54}$Cr double spike was prepared from enriched $^{50}$Cr (96.5%; oxide form) and $^{54}$Cr (95.5%; metal form), purchased from CortecNet (Voisins-le-Bretonneux, France). The single spikes were separately digested in 6 M HCl at ca. 120 °C on a hotplate ($^{54}$Cr) and in 15 M HNO$_3$ in a Parr Bomb at ca. 160 °C for several days ($^{50}$Cr), respectively. After digestion, the $^{54}$Cr was evaporated and re-dissolved in 15 M HNO$_3$. The optimal spike composition, *i.e.*, 0.58:0.42 $^{50}$Cr:$^{54}$Cr, was calculated using the double spike toolbox (58). The two digested single spikes were mixed accordingly, and the double spike was further diluted with 2M HNO$_3$ to a concentration of about 38 µg/ml. The $^{50}$Cr-$^{54}$Cr double spike was calibrated relative to the reference material NIST SRM 979 for spike:sample mixing ratios of 0.13:0.87 to 0.39:0.61. The optimal mixing ratio according to ref. (58) is 0.28:0.72.

*Sample preparation and purification*

Approximately 30 mg of sample powder was dissolved in a concentrated HCl-HF-HNO$_3$ mixture (1 mL, 0.5 mL, 0.2 mL, respectively) at 130°C for 48 hours and afterwards evaporated. Subsequently, the samples were re-dissolved in concentrated HNO$_3$-HF (2 mL, 0.2 mL, respectively) and placed in Parr Bombs under pressurised steel jackets at 190 °C in an oven for 72 hours, to dissolve refractory chromite. No residues were observed upon visual inspection of the digested samples. An adequate amount of $^{50}$Cr-$^{54}$Cr double spike was added to the samples following dissolution, and sample and double spike were allowed to equilibrate in closed PFA sample vials at around 120 °C overnight. Chemical separation of Cr from matrix elements (*e.g.*, Mg, Ca and Fe), and particularly from elements causing isobaric interferences on the Cr masses during mass spectrometry (*i.e.,* Fe, Ti, V), was achieved using a two-step chromatography procedure with AG50W-X8 (200-400 mesh) cation exchange resin, adapted from ref. (59). The first cation exchange column, custom-made from heat-shrinkable Teflon, accommodated 1.1 ml resin, and was conditioned with 0.5 M HCl. Evaporated samples were re-dissolved and loaded in 1.1 ml of 0.55 M HCl, and Cr was collected immediately upon sample loading. Further elution of Cr was achieved in 4 ml 1M HCl. Major matrix elements, such as Mg, Ca and Fe remained sorbed to the resin, quantitatively removing them from the Cr-bearing eluate during this step. As minor amounts of Al, Ti and V co-elute with Cr, a second cation-exchange column was used to further purify the samples. The evaporated Cr fraction was re-dissolved in 2.1 ml 0.8 M HNO$_3$ and loaded onto custom-made Teflon micro-columns, accommodating 300 µl of cation-exchange resin, which was conditioned with MilliQ de-ionised H$_2$O. Matrix elements were removed with 2.5 ml 0.8 M HF and 6 ml 1 M HCl before Cr was collected in 3 ml 6 M HCl. Recovery of Cr from this procedure was between 60-90%. Although stable isotope fractionation is caused by incomplete elution of Cr (42), this, along with other sources of mass bias during analysis, is corrected for by the double spike reduction. Following chromatographic purification, the samples were evaporated under concentrated $HNO_3$, diluted to a concentration of 1 ppm Cr, and dissolved in 2% v/v (0.317 M) $HNO_3$ for isotope analysis.

*Mass spectrometry and data reduction*

The purified Cr samples were introduced, *via* a Cyclonic-Scott Double Pass spray chamber into a Thermo Scientific Neptune Plus Multi-Collector Inductively-Coupled Plasma Mass Spectrometer housed at the Institut de Physique du Globe de Paris, Paris, France. Medium resolution mode, which permits resolution of polyatomic interferences (*e.g.*, $^{40}Ar^{14}N^+$ on $^{54}Cr^+$ and $^{40}Ar^{16}O^+$ on $^{56}Fe^+$) was employed, and analyses were performed on peak shoulders. Isobaric interferences from $^{54}Fe^+$ on $^{54}Cr^+$, and $^{50}Ti^+$ and $^{50}V^+$ on $^{50}Cr^+$ were monitored on $^{56}Fe^+$, $^{49}Ti^+$ and $^{51}V^+$, respectively, and corrected for using the exponential mass fractionation law. Samples were bracketed by the NIST SRM 979 reference material (standard-sample-sample-standard) and analysed between 2 and 3 times each, with each analysis consisting of 100 cycles of 4.194s integration time. Instrumental mass bias was corrected for by the $^{50}Cr$-$^{54}Cr$ double spike, following the iterative solution of ref. (60); assuming exponential instead of linear fractionation. The isotopic ratio is reported in delta notation:

$$\delta^{53}Cr\ (‰) = \left(\frac{(^{53}Cr/^{52}Cr)_{sample}}{(^{53}Cr/^{52}Cr)_{NIST\ SRM\ 979}} - 1\right) \times 1000. \quad (7)$$

The intermediate measurement precision was estimated by calculating the pooled standard deviation ($s_p$) of the sample data set:

$$s_p = \sqrt{\frac{\sum_{i=1}^{k}(n_i-1)s_i^2}{\sum_{i=1}^{k}(n_i-1)}}, \quad (8)$$

where $n_i$ is the number of repeated measurements of each sample *i* and $s_i^2$ their variance. The resulting $2s_p$ of ± 0.03 ‰ (k = 35) is identical to the 2 standard deviation of repeated analyses of the USGS geological reference material BHVO-2 (-0.104 ± 0.03 ‰; n = 4). The $\delta^{53}Cr$ value of the BHVO-2 is further in agreement with literature data (61–63), assuring accuracy of the analyses.

**Acknowledgements**

We thank Justin Simon, an anonymous reviewer, and the editor for their careful and thorough input that vastly improved the end product. We are grateful to CAPTEM for provision of lunar samples. P.A.S. and F.M. acknowledge funding from the European Research Council under the H2020 framework program/ERC grant agreement #637503 (Pristine) as well as financial support of the UnivEarthS Labex program at Sorbonne Paris Cité (ANR-10-LABX-0023 and ANR-11-IDEX-0005-02), and the ANR through a chaire d'excellence Sorbonne Paris Cité. Parts of the analytical facilities were supported by IPGP multidisciplinary program PARI, and by Paris-IdF region SESAME Grant No. 12015908. P.A.S. is indebted to Hugh O'Neill, Bruce Fegley, Marc Norman and Ryuki Hyodo for discussions. P.A.S. conceived the project, performed the analyses, interpreted the data and wrote the paper. F.M. conceived the project, helped to interpret the data and write the paper. K.v.Z. developed the analytical procedure and performed some of the analyses. The authors declare no competing interests. This open access article is distributed under Creative Commons Attribution-Non Commercial-No Derivatives License 4.0 (CC BY-NC-ND). All data is available in the main text or in the supporting information online at www.pnas.org/lookup/suppl/doi:10.1073/pnas.1809060115/-/DCSupplemental.